\documentclass[aps,prd,onecolumn,groupedaddress,showpacs,nofootinbib,amssymb]{revtex4}
\usepackage[T1]{fontenc}
\usepackage[latin1]{inputenc}
\usepackage{graphicx}
\usepackage[english]{babel}
\usepackage{amsmath}
\usepackage{amssymb}
\usepackage{amsfonts}
\usepackage{colordvi}
\usepackage{color}

\begin{document}

\title{Topological invariant quintessence}
\author{ Mariafelicia De Laurentis\footnote{e-mail address: mfdelaurentis@tspu.edu.ru}}
\affiliation{Tomsk State Pedagogical University, Tomsk, ul. Kievskaya 60, 634061 Russian Federation}
\date{\today}

\begin{abstract}
The issues of quintessence and cosmic acceleration can be discussed in
the framework of $F(R, {\cal G})$ theories of gravity where $R$ is the Ricci curvature scalar and ${\cal G}$ is the Gauss-Bonnet topological invariant.
It is possible to show that such an approach exhausts all the curvature content related to the Riemann tensor giving rise to a fully geometric approach to dark energy.
 \end{abstract}
 
 \maketitle

\section{Introduction}
The apparently accelerated behavior of cosmic Hubble fluid poses the problem to explain the source of such a dynamics outside the standard matter-energy perfect fluids. Besides dark matter, dark energy is one of the main puzzle of modern cosmology which requires new paradigm to be explained. In fact, the standard $\Lambda$CDM model, also being a relevant snapshot of the presently observed universe, suffers  several shortcomings as soon as one wants to describe in a self-consistent way the whole cosmological evolution from early to late epochs.  

The search for a solution to this issue lies on two main approaches: from one side, dark energy could be related to new material ingredients as scalar fields, on the other hand, it could be related to  extensions of General Relativity that behave differently with respect to  Einstein's theory at infrared scales.

Recently, a new class of modified gravity theories, whose action contains  general functions of the Ricci scalar $R$  and the Gauss-Bonnet invariant ${\cal G}$, {\it i.e.}  $F(R, {\cal G})$, gave rise to a lot of interests in the context of the Extended Theories of Gravity\cite{PhysRepnostro,OdintsovPR,5,6,Mauro,faraoni,9,10,libri,libroSV,libroSF}. These models  are  appealing for several reasons \cite{F(GR)-gravity,17,18,20,21,23,24,28,32,ester}. First,  they not to introduce gravitational ghosts in the spin-$2$ sector, but could produce ghost degrees of freedom in an empty anisotropic universe, as in the case of Kasner-type background \cite{kasner}. It is worth noticing that these degrees
of freedom are totally absent on Friedmann-Robertson-Walker cosmological backgrounds. Furthermore, they are able to describe the present acceleration of the universe, as well as,  the transitions from acceleration to deceleration phases going back with the redshift \cite{OdintsovPR}. Besides, it is possible to reproduce several cosmological solutions using also the Noether Symmetries approach  \cite{diego,myrz,diego3,Felice_malo,fGBnoether}. Models that take into account the stability of cosmological solutions are used to derive self-consistent  energy conditions and perturbation theory \cite{defeliceper,darabi}. Finally, spherically  symmetric solutions \cite{zerb},  as well as  Parametrized Post-Newtonian expansions has been found \cite{antonio}. 

Besides the already achieved results, some fundamental issues are related to the presence of both the  fields $R$ and ${\cal G}$. In any formulation of quantum field theory on curved spaces, such fields appear in order to regularize the effective action \cite{birrel}. In other words, together with invariants constructed with  $R_{\mu\nu}$ and $\Box R$, these fields give contributions  to the trace anomaly and then allow to deal with gravity, at least at one-loop level, under the same standard of matter fields \cite{PhysRepnostro}. From the point of view of gravitation, it seems that considering $R$ and ${\cal G}$ takes into account  all the degrees of freedom related to the curvature invariants and then shortcomings related to ghosts and anomalies (at infrared and ultraviolet levels) can be removed.  However, also this is an effective description but it could be  a relevant path towards quantum gravity.

Similar considerations work also in cosmology. Several shortcomings related to $F(R)$ gravity could be addressed considering $F(R,{\cal G})$ since the whole budget of curvature degrees of freedom are taken into account. In this case, at extremely large  infrared scales. 

In this paper, following the track of $F(R)$ gravity cosmology \cite{quintessence}, we want to take into account  the possibility that dark energy behavior can be addressed by $F(R,{\cal G})$. Some papers already deal with this problem: Here, we want to point out how these geometric double-field models could enhance the early $F(R)$ taking care of the other curvature degrees of freedom. 

In Sec.\ref{due}, we derive the $F(R,{\cal G})$ field equations. Cosmology related  to $F(R,{\cal G})$ is considered in  Sec.\ref{tre} while topological invariant quintessence is discussed in Sec. \ref{quattro}. Examples of exact solutions are also given. Conclusions are drawn in Sec. \ref{cinque}.

\section{$F(R,\cal{G})$ gravity} \label{due}

A general action including the Ricci scalar $R$ and the  Gauss-Bonnet invariant $\cal G$ is 
\begin{equation}
{\cal A}=\frac{1}{2\kappa}\int d^4x \sqrt{-g}\left[F(R,{\cal G})+{\cal L}_M\right]\,,
   \label{action}
\end{equation}
where 
${\cal L}_M$ is the matter Lagrangian.   The Gauss-Bonnet invariant is defined as
\begin{equation}
{\cal G}\equiv
R^2-4R_{\mu\nu}R^{\mu\nu}+R_{\mu\nu\rho\sigma}
R^{\mu\nu\rho\sigma}\,,
   \label{GBinvariant}
\end{equation}
so, in principle, the action (\ref{action}) contains all the possible curvature invariants that can be derived starting from the Riemann tensor.
We are using physical units $c= k_B=\hbar=1$.
The variation of action (\ref{action}) with respect to the metric provides the following gravitational field equations
\begin{equation}\label{eom}
R_{\mu \nu}-\tfrac{1}{2} g_{\mu \nu}R=\kappa\,T^{(mat)}_{\mu \nu}+
T^{(\cal{GB})}_{\mu \nu},
\end{equation}
where $\kappa=8\pi G$,  $T^{(mat)}_{\mu \nu}=diag\left(-\rho^{(m)},p^{(m)},p^{(m)},p^{(m)}\right)$ is the stress energy tensor describing the ordinary matter, with $\rho^{(m)}$ and $p^{(m)}$ being respectively, the matter energy density and pressure. The extra term  $T_{\mu\nu}^{(\cal{GB})}$, containing all extra curvature terms, is defined as 
\begin{widetext}
\begin{eqnarray}
&&T^{(\cal{GB})}_{\mu \nu}=\nabla_\mu \nabla_\nu  \frac{\partial F(R,{\cal G})}{\partial R}-g_{\mu \nu}\, \Box\,  \frac{\partial F(R,{\cal G})}{\partial R}+2R \nabla_\mu \nabla_\nu  \frac{\partial F(R,{\cal G})}{\partial {\cal G}}-2g_{\mu \nu} R\, \Box\,  \frac{\partial F(R,{\cal G})}{\partial {\cal G}}-4R_\mu^{~\lambda} \nabla_\lambda \nabla_\nu  \frac{\partial F(R,{\cal G})}{\partial {\cal G}}\nonumber\\&&\nonumber\\&& -4R_\nu^{~\lambda} \nabla_\lambda \nabla_\mu  \frac{\partial F(R,{\cal G})}{\partial {\cal G}}
+4R_{\mu \nu} \Box  \frac{\partial F(R,{\cal G})}{\partial {\cal G}}+4 g_{\mu \nu} R^{\alpha \beta} \nabla_\alpha \nabla_\beta  \frac{\partial F(R,{\cal G})}{\partial {\cal G}}+4R_{\mu \alpha \beta \nu} \nabla^\alpha \nabla^\beta  \frac{\partial F(R,{\cal G})}{\partial {\cal G}}\nonumber\\&&\nonumber\\&&-\frac{1}{2}\,g_{\mu \nu} \left[ R \frac{\partial F(R,{\cal G})}{\partial R} + {\cal G}  \frac{\partial F(R,{\cal G})}{\partial {\cal G}}- F(R,{\cal G})\right] +\left(1- \frac{\partial F(R,{\cal G})}{\partial R}\right)\,\left(R_{\mu \nu}-\frac{1}{2} g_{\mu \nu}R\right)\,,
\label{effective-energy-momentum}
\end{eqnarray}
\end{widetext}
where $\nabla$ is the  covariant derivative  and $\Box$ is the d'Alembert operator.  
%The trace equation is
%
% \begin{equation}\label{8g33}
%                    - 2 F(R, {\cal G}) +  R \frac{\partial F(R,{\cal G})}{\partial R}  + 3 \nabla^2 \frac{\partial F(R,{\cal G})}{\partial R}  + 2 {\cal G}  \frac{\partial F(R,{\cal G})}{\partial {\cal G}}  + 2 R \nabla^2  \frac{\partial F(R,{\cal G})}{\partial {\cal G}} - 4 R_{\rho \sigma}
%                    \nabla^\rho \nabla^\sigma  \frac{\partial F(R,{\cal G})}{\partial {\cal G}} = 0.
%                \end{equation}
 Einstein's gravity is immediately recovered for  $F(R,{\cal G}) = R$.
 %It is interesting to note that the following terms that appear in the definition the effective of energy momentum tensor (\ref{effective-energy-momentum}) is a sort of potential 
%

In the following, for the sake of   simplicity, we denote by 

\begin{equation}
  \label{eq:def1}
  F_{R}\equiv\frac{\partial F(R,{\cal G})}{\partial R}\,,\qquad F_{\cal G}\equiv \frac{\partial F(R,{\cal G})}{\partial {\cal G}}\,, 
\end{equation}
the partial derivatives with respect to $R$ and $\cal G$. From the above form  (\ref{effective-energy-momentum}), it is clear that the form of $F(R,{\cal G})$ determine the dynamical behavior of the theory. In  particular, the term 
 \begin{equation}
W(R,{\cal G})\equiv R \frac{\partial F(R,{\cal G})}{\partial R} + {\cal G}  \frac{\partial F(R,{\cal G})}{\partial {\cal G}}- F(R,{\cal G})\,,
\end{equation}
can be considered as an effective double-field potential for the theory.  Clearly, $R$ and $ \cal G$ act as two different scalar fields whose regimes can lead different phases of the cosmological evolution.

Below, we are going to derive the cosmological equations in order to discuss the quintessential behavior.

\section{$F(R,\cal G)$ cosmology}\label{tre}
In order to deal with cosmology, 
let us consider the  Friedman-Robertson-Walker  metric
\begin{equation}
ds^{2}=-dt^{2}+a^{2}(t)(d{x}^{2}+d{y}^{2}+d{z}^{2})\,,
\label{metric}
\end{equation}
where $a(t)$ is the scale factor. We assume flat models.  The above field equations reduce to  

%\begin{eqnarray}\label{FRWa}

\begin{eqnarray}\label{FRWa}
\left(\frac{{\ddot a}a-{\dot a}^2}{a^2}\right)F_R&=&
-\frac{\kappa}{2}\left(p^{(m)}+\rho^{(m)}\right)+
\frac{1}{
2}\left[\left(\frac{\dot a}{a} \right){\dot F}_R-\ddot{F}_R+4 \left(\frac{\dot a}{a}\right)^3\dot{F}_{\cal G}-8 \left(\frac{\dot a}{a}\right) \left(\frac{{\ddot a}a-{\dot a}^2}{a^2}\right)\dot{F}_{\cal G}-4\left(\frac{\dot a}{a}\right)^2\ddot{F}_{\cal G}\right],\nonumber\\
\\ 
F_R \left(\frac{\dot a}{a}\right)^2&=&
{\kappa\over 3}\rho^{(m)}+
\frac{1}{6}\left[R\,F_R-F(R,{\cal G})-6\left(\frac{\dot a}{a}\right)\dot{F}_R+{\cal G}F_{\cal G}-24\left(\frac{\dot a}{a}\right)^3\dot{F}_{\cal G}\right]\nonumber\,,
\end{eqnarray}
where 
$\rho^{(m)}$ and $p^{(m)}$ are the energy density and pressure of ordinary matter, respectively, and 
the overdot denotes a derivative with respect to the cosmic time  $t$.

In addition, the Hubble parameter is ${\displaystyle H=\frac{\dot a}{a}}$ and the fields $R$ and $\cal G$ assume the forms
\begin{eqnarray}
R &=& 6 \left[\frac{\ddot a}{a}+\left(\frac{\dot a}{a}\right)^2\right]= 6 \left(2H^{2}+\dot H \right)\,,
\label{eq:R} \\
{\cal G} &=&\frac{24 {\dot a}^2 {\ddot a}}{a^3}= 24H^{2} \left( H^{2}+\dot H \right)\,. \label{eq:G}
\end{eqnarray}
The above Friedmann equations (\ref{FRWa}) become

\begin{eqnarray}\label{FRWH}
F_R\dot{H}&=&
-\frac{\kappa}{2}\left(p^{(m)}+\rho^{(m)}\right)+
\frac{1}{2}\left[H \dot{F}_R-\ddot{F}_R+4H^3\dot{F}_{\cal G}-8H\dot{H}\dot{F}_{\cal G}-4H^2\ddot{F}_{\cal G}\right],\nonumber\\
\\
F_R H^2&=&
{\kappa\over 3}\rho^{(m)}+
\frac{1}{6}\left[F_RR-F(R,{\cal G})-6H\dot{F}_R+{\cal G}F_{\cal G}-24H^3\dot{F}_{\cal G}\right]\nonumber.
\end{eqnarray}
and can  be rewritten as 
\begin{equation}
\rho_{(tot)}=\frac{3}{\kappa}H^{2}\,,
\quad
p_{(tot)}=-\frac{1}{\kappa} \left( 2\dot H+3H^{2} \right)\,,
\label{GutenTag}
\end{equation}
where $\rho_{(tot)}$ and $p_{(tot)}$ are
the energy density and pressure due to the introduction of Gassu-Bonnet term, respectively, defined by

\begin{eqnarray}
\rho_{(tot)} &= &\frac{1}{F_R}\left[\rho^{(m)}+\frac{1}{2\kappa}\left(R F_R-F(R,{\cal G})-6H\dot{F}_R+{\cal G}F_{\cal G}-24H^3\dot{F}_{\cal G}\right)\right]\,, \label{eq:rho-eff-1}
\\
\nonumber\\
p_{(tot)} &=& \frac{1}{F_R}\left\{p^{(m)}+\frac{1}{\kappa}\left[2H\dot{F}_R+\ddot{F}_{R}+8H^3\dot{F}_{\cal G}+8H\dot{H}\dot{F}_{\cal G}+4H^{2}\ddot{F}_{\cal G}-\frac{1}{2}\left(R F_R+{\cal G} F_{\cal G}-F(R,{\cal G}) \right)\right]\right\}
\,,
\label{eq:p-eff-1}
\end{eqnarray}
where the further contributions to pressure and energy density are clearly given by $R$ and Gauss-Bonnet contributions. In what follows, we will show that quintessential behavior can be achieved thanks to geometrical terms. 

\section{Topological Invariant quintessence}\label{quattro}
Astrophysical evidences show that the universe is now 
undergoing a phase of accelerating behavior right \cite{perlmutter}. 
It is clear that inflation solves a lot of issues in the early universe,
but, the presence of accelerated expansion at present epoch creates problems since the source of such a further acceleration is highly questionable.  The simplest way to solve this puzzle is that  we can describe accelerating behavior by adding a cosmological constant to the Einstein field equations. 
The problem with this approach is, that such a constant, should be extremely small compared to quantum gravity scales so we need a mechanism capable of evolving   the energy density and pressure associated with the cosmological constant.
Secondly, the cosmological constant starts dominating the energy density of the universe
very quickly and if we want it to dominate right now, it should have been tremendously small in the early universe. The {\it quintessence} problem, as it is called, actually consists
of two related problems. The  fine-tuning: why is the cosmological constant starting to dominate the energy density  at present epoch?
The magnitude fine-tuning: why is the energy density of the cosmological constant so
small and comparable with matter density?
Solving one of these questions often means that the other question is automatically 
answered. Several  cosmologists are reluctant to insert  the cosmological constant by hand, since it involves setting its value just right to solve the fine-tuning problems. It is also possible to invoke dynamical mechanisms which are hoped to cope in a natural way with the smallness and timing problem. Although a lot of  models have been built in order to address the problem without assuming a cosmological constant term, there is no general consensus regarding the final solution. 
In particular, it is possible to find models which explicitly violate energy  conditions but satisfy no-hair theorem requests assuring a cosmological constant behavior.  Precisely, this fact happens if Extended  Theories of Gravity are involved and matter is in the form of scalar fields, besides the ordinary perfect fluid matter \cite{quartic}. In this context, a fundamental issue is to select the classes of gravitational theories and the conditions which "naturally" allow to recover an effective cosmological constant \cite{effective}.

With these considerations in mind, 
 an  accelerated behavior is achieved if

\begin{equation}
\label{20} \rho_{(tot)}+ 3p_{(tot)}< 0\,,
\end{equation}
that means, in the specific case of $F(R,\cal G)$,
\begin{eqnarray}
&&\frac{1}{F_R}\left(\rho^{(m)}+3p^{(m}\right)+
 \frac{{\cal G}F_{\cal G}-F(R,{\cal G})+R F_R-24 H^3 {\dot F}_{\cal G}-6 H {\dot F}_R}{2 \kappa F_R }\nonumber\\
&&+\frac{3 \left[\frac{1}{2} \left(F(R,{\cal G})-{\cal G}F_{\cal G}- R F_R \right)+4 H^2 {\ddot F}_{\cal G}+8 H^3 {\dot F}_{\cal G}+8 H {\dot F}_{\cal G} {\dot H}+2 H {\dot F}_R {\ddot F}_R\right]}{\kappa F_R}<0\,.
\end{eqnarray}
We can write also
\begin{equation}
\label{21} \rho_{(\cal GB)}> \frac{1}{3}\rho_{(tot)}\,,
\end{equation}
assuming that all matter components have non-negative pressure.
In other words, since standard matter does not violate energy conditions,
 acceleration condition depends on the relation
\begin{eqnarray}\label{eq:NEC}
\nonumber
\rho_{(\cal{GB})}+3 p_{(\cal{GB})}&=& \frac{F(R,{\cal G})-{\cal G}F_{\cal G}-R F_R+3 H \left[4 H {\ddot F}_{\cal G}+4 {\dot F}_{\cal G} \left(H^2+2 {\dot H}\right)+{\dot F}_R \left(2 {\ddot F}_R-1\right)\right]}{\kappa F_R}\,,
\end{eqnarray}
which has to be compared with matter contribution. However, we can define
\begin{equation}
 \label{23}
 w_{(GB)}=\frac{p_{(\cal{GB})}}{\rho_{(\cal{GB})}}\,,
% \qquad -1\leq w_{(\cal{GB})}<0\,.
 \end{equation}
 which is the equation of state coming from geometry that explicitly is 
 \begin{eqnarray}   
w_{(\cal{GB})}= \frac{{\cal G} F_{\cal G}
   +R F_R-F(R,{\cal G})-4 H \left[2 H {\ddot F}_{\cal G}+4 {\dot F}_{\cal G} \left(H^2+{\dot H}\right)+{\dot F}_R {\ddot F}_R\right]}{F(R,{\cal G})+24 H^3 {\dot F}_{\cal G}-{\cal G} F_{\cal G} +6 H {\dot F}_R-R F_R }\,.
 %  +\frac{p^{(m)}}{\rho^ {(m)}} 
     \end{eqnarray}
Clearly, for standard matter is 
\begin{equation}
0\leq w_{(m)}<1\,,
 \end{equation} 
in  the so-called Zeldovich interval where all the forms of ordinary matter fluid are
enclosed ($w_{(m)}=0$ gives {\it dust}, $w_{(m)}=\frac{1}{3}$ is {\it radiation}, $w_{(m)}=1$ is {\it stiff matter}).
 Then quintessence means
\begin{equation}
 \label{24}
-1\leq w_{(\cal{GB})}<0\,,
 \end{equation}   
 while the phantom behavior is achieved for 
 \begin{equation}
 \label{25}
w_{(\cal{GB})}<-1\,.
 \end{equation}   
 The form of $F(R,{\cal G})$ is the main ingredient to obtain  the various behaviors. We will call the condition (\ref{24}) {\it topological invariant quintessence}.
 
Clearly, the possibility to achieve such behaviors are infinite. A criterion to obtain   viable $f(R,{\cal G})$ is to consider the so called {\it Noether Symmetry Approach}. In  \cite{fGBnoether}, the  form 
\begin{eqnarray}\label{FRG}
F({R,\cal G})= F_0 R^{n} {\cal G}^{1-n}.
  \end{eqnarray}
 has been obtained, where $n$ is a real number. For power law
 solutions of   the form
 \begin{eqnarray}
a(t)=a_0t^s\,,
  \end{eqnarray}
  where the relations
  \begin{equation}
   n=\frac{1+s}{2}\;\;\;\;\;\; \;\; n=\frac{1}{1+2s(s-1)}-2s\;,
\end{equation} 
hold. Interesting cases for quintessence are obtained for $n\neq 1$ and $s\geq 1$ (accelerated behaviour). 
Converting the scale factor and the cosmic time in terms of the red shift $z$, we have
\begin{eqnarray}
\frac{a(t)}{a_0}=\frac{1}{1+z} \;\;\;\; {\text{and}}  \;\;\;\;\ t=t_0(1+z)^{-\frac{1}{s}}\, 
\end{eqnarray}
and then the above {\it Gauss-Bonnet source} (\ref{eq:NEC}) of cosmological equations for any  $n$ and $s$ is 
\begin{eqnarray}
\rho_{\cal GB}+3p_{\cal GB}&=&\frac{3 (n-1)}{k (s-1) s}\left\{-15+s\left[ 45-30s-\left[6+4n(1-3s)^2(2s-1)+(5s+3s^2(8s-15))\right](1+z)^{\frac{2}{s}}-\right.\right.\nonumber\\&&\left.\left. 15\times 4^{2-n} (n-1) n (s-1)^2 (3 s-1) (z+1)^{\frac{2}{s}} \left[(2 s^2-s) (z+1)^{\frac{2}{s}}\right]^{n-1} \left[(s-1) s^3 (z+1)^{\frac{4}{s}}\right]^{1-n}-\right.\right.\nonumber\\&&\left.\left. \frac{4^{2-n}}{s^2}(n^2-n)(3s-1)\left[4n(1-3s)^2-6-9s-15s^2\right] \left[(2 s^2-s) (z+1)^{2/s}\right]^{n-1}\times\right.\right.\nonumber\\&&\left.\left. \left[(s-1) s^3 (z+1)^{4/s}\right]^{2-n}  \right]\right\}\,.
\label{prGB}
\end{eqnarray}
Immediately, the equation of state becomes
\begin{eqnarray}\label{wz}
w_{(\cal GB)}&=&\frac{1}{6} \left\{ 6-\frac{17}{s}+\frac{14}{3s-1}+\frac{1}{2 s-1}\left[ \frac{2 (1-2 s)^2 \left(2 n (1-3 s)^2-3\right)}{s^2 (3 s-1)}+ \frac{15 (1-2 s)^2 (s-1) (z+1)^{-2/s}}{s^3 (3 s-1)}+\right.\right.\nonumber\\&&\left.\left.      \frac{4^{2-n} (n-1) n (s-1) \left[4 n (1-3 s)^2-3 (5 s^2+3s+2)\right]\left[(2 s^2-s) (z+1)^{\frac{2}{s}}\right]^n \left[(s-1) s^3
   (z+1)^{\frac{4}{s}}\right]^{1-n}}{s^2}+\right.\right.\nonumber\\&&\left.\left. 15\times 4^{2-n} (n-1) n (s-1)^3 (z+1)^{\frac{2}{s}} \left[ (2 s^2-s) (z+1)^{\frac{2}{s}}\right]^n \left[(s-1) s^3 (z+1)^{\frac{4}{s}}\right]^{-n}
 \right]     \right\}\,,
\end{eqnarray}
that can be recast in the general form 
\begin{equation}\label{wzGB}
\boxed{ w_{(\cal GB)}=w_0+w_1(1+z)^{-\frac{2}{s}}+w_2 (1+z)^{\frac{4-2n}{s}}+w_3 (1+z)^{\frac{2-2n}{s}}}
\end{equation}   
where, in principle, all the cosmological behaviors can be achieved according to the values of $s$ and $n$. 

Immediately, $\Lambda$CDM can be recovered for $w_0=-1$ and $w_1=w_2=w_3=0$. For $w_1\neq 0$ the equation of state is evolving according to the redshift $z$.

For  $n=2$ and $s=3$,  we have that $a(t)$ evolves as a power law and

\begin{equation}
\label{sol1r3p}
\rho_{(\cal GB)}+3p_{(\cal GB)}= k_0(z+1)^{2/3}+k_1\,,
\end{equation}
and

\begin{equation}\label{wzGB1}
 w_{(\cal GB)}=w_0+w_1(1+z)^{-\frac{2}{3}}\,.
\end{equation}  
The constant $k_{0,1}$ are negatively defined addressing immediately the quintessence issue of accelerating behavior.
Clearly, the form of $F(R,\cal G)$ rules the cosmological behavior.

\section{Conclusions}\label{cinque}

In this paper, we have shown that the {\it quintessence paradigm} can be recovered in the framework of $F(R,{\cal G)}$ theories of gravity. In other words, as it is possible for inflationary models, we can ask for a sort of {\it topological  invariant quintessence} which can be recovered by taking into account curvature and topological invariants. Inserting also $\cal G$ into dynamics allows, in some sense, to consider all the curvature budget coming from the Riemann tensor. This fact improve the curvature approach to dark energy related to adopting $F(R)$ gravity and is strictly related to effective theories since the Gauss-Bonnet invariant comes out from defining quantum fields in curved spacetimes. In a forthcoming paper, we will confront $F(R,\cal G)$ with respect to cosmography in order to derive more physically motivated models.

\end{document}